\documentclass{vgtc}                          % final (conference style)
\graphicspath{{figures/}{pictures/}{images/}{./}} % where to search for the images

\usepackage{times}                     % we use Times as the main font
\usepackage{tabu}                      % only used for the table example
\usepackage{booktabs}                  % only used for the table example
\usepackage{lipsum}                    % used to generate placeholder text
\usepackage{mwe}                       % used to generate placeholder figures
\usepackage{amsmath}
\usepackage{cite}

\usepackage{mathptmx}                  % use matching math font

\onlineid{0}

\vgtccategory{Research}

\vgtcinsertpkg

\newcommand{\papername}{PlanePivoting\xspace}
\title{\papername: Exploration and Optimization of Gaze-Mouse Cursor Alignment for Spatial Object Translation}
\author{Jinwook Kim\thanks{e-mail: jinwook.kim31@kaist.ac.kr}\\ %
        \scriptsize KAIST %
\and Sangmin Park\thanks{e-mail: psmdc0714@kaist.ac.kr}\\ %
        \scriptsize KAIST %
\and Jihyeon Lee\thanks{e-mail: jihyeon@kaist.ac.kr}\\ %
        \scriptsize KAIST %
\and Sang Ho Yoon\thanks{e-mail: sangho@kaist.ac.kr}\\ %
        \scriptsize KAIST %
\and Jeongmi Lee\thanks{e-mail: jeongmi@kaist.ac.kr}\\ %
        \scriptsize KAIST }

\teaser{
  \centering
  \includegraphics[width=0.95\linewidth]{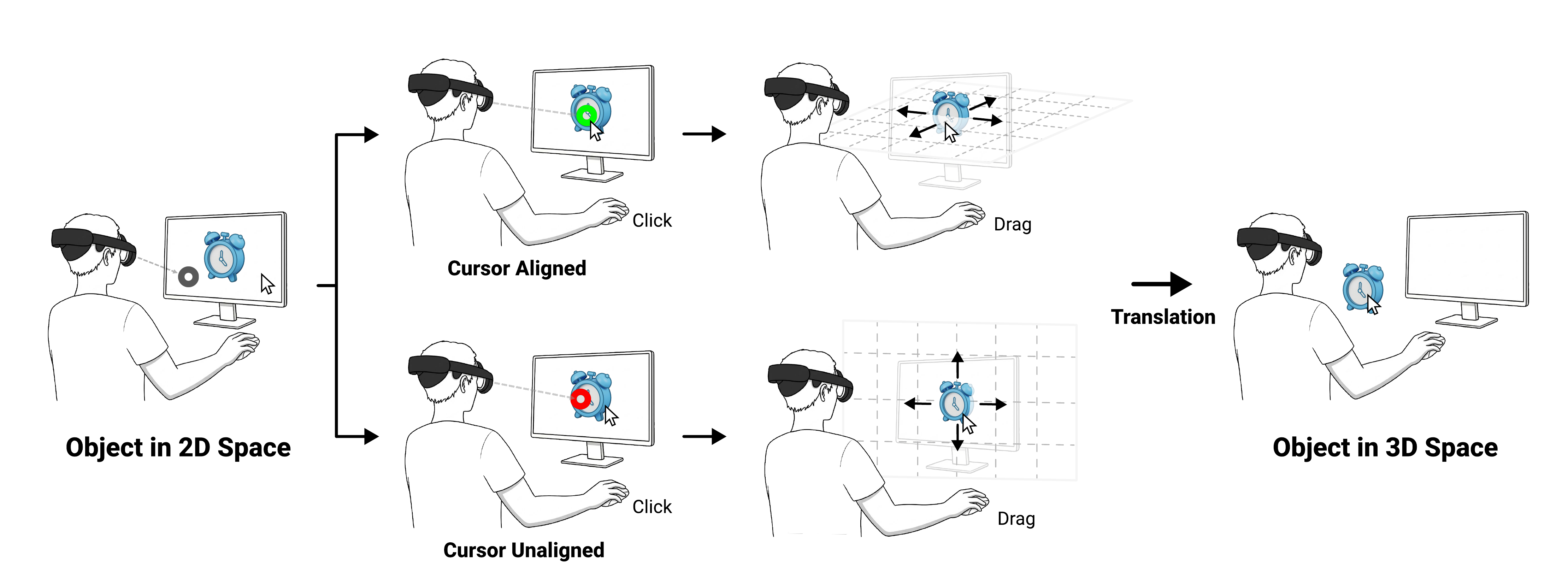}
  \caption{Illustration of \papername (overlapXZ), a technique that determines the 3D translation plane based on the interaction between the gaze and mouse cursors. When users attempt to select the object (3D Clock Model), the system detects whether the gaze and mouse cursors overlap or are separated; this state determines whether the user drags the object in the horizontal or vertical plane. After placing the object at the target location, the user releases the mouse to finalize the placement.}
  \label{fig:teaser}
}

\abstract{
As XR matures into a ubiquitous computing platform, the disconnect between 2D and 3D input modalities remains a critical barrier to seamless workflow. Frequent transitions between the mouse for 2D precision and hand gestures for 3D manipulation induce significant physical fatigue and cognitive load. To address this, we introduce \papername, a multimodal interaction technique that extends standard mouse input into 3D space by leveraging gaze-mouse alignment. This technique dynamically modulates the translation plane based on the spatial overlap between the gaze and mouse cursor, eliminating the need for physical input modality switching. To systematically explore the foundational design space of gaze-mouse coordination and optimize key variables, we conducted a user study comparing \papername with a standard 3D Gizmo interface across two translation mapping profiles and two gaze cursor apertures. Results demonstrate that \papername outperforms the Gizmo on efficiency metrics while maintaining comparable precision and yielding higher subjective satisfaction. This study demonstrates the potential of gaze-mouse alignment for efficient spatial manipulation between 2D and 3D environments.
}


\begin{document}



%%%%%%%%%%%%%%%%%%%%%%%%%%%%%%%%%%%%%%%%%%%%%%%%%%%%%%%%%%%%%%%%%%%%%%%%%%%%%%%%%%%%%%%%%%%%%%%%%%%%%%%%%%%%%%
\firstsection{Introduction}
\maketitle
As Extended Reality (XR) technologies mature, they are increasingly adopted for productivity tasks, evolving into spatial computing workstations and extended desktop environments~\cite{gonzalez2024guidelines}. In these seated XR scenarios, users still rely on the familiar keyboard and mouse for high-precision 2D tasks. However, as the workspace expands beyond physical monitors into the surrounding 3D space, current input modalities create a disjointed user experience~\cite{wang2022design, brudy2019cross}. The mouse is inherently suboptimal for volumetric manipulations in XR (i.e., translation, rotation) due to its limited degrees of freedom~\cite{tutuncu2025handover}. Conversely, ray-based interactions (e.g., controllers, Gaze+Pinch~\cite{pfeuffer2017gaze}) lack the precision required for traditional 2D interfaces, particularly when targeting small UI elements~\cite{lee2025facilitating}. Consequently, users are compelled to switch interaction techniques frequently as they transition between 2D flat display environments and volumetric XR spaces~\cite{cools2025comparison, gonzalez2024guidelines}.

Recent studies have explored novel interaction techniques to facilitate seamless transitions between these heterogeneous environments~\cite{tutuncu2025handover, rau2025traversing}. For instance, Rau et al.~\cite{rau2025traversing} proposed cross-reality techniques that allow users to transfer 3D content from a 2D desktop screen to an XR space using gestural transitions. While these approaches effectively blend 2D and 3D modalities, they still necessitate frequent physical transitions between grabbing the mouse and performing bare-hand gestures. In a seated productivity context where a user’s hand is already resting on the mouse, forcing a shift to mid-air gestures to frequently arrange 3D assets across 2D and 3D spaces or position spatial widgets can disrupt workflow continuity and induce significant physical fatigue and cognitive load~\cite{cools2025comparison}. Therefore, it is crucial to extend standard mouse interaction into 3D space, thereby eliminating the need to switch input methods.

%Recent cross-reality research highlights the importance of minimizing interaction costs during input transitions. Cools et al.~\cite{cools2025comparison} demonstrated that while the mouse ensures the fastest performance due to familiarity, users prefer the flexibility to strategically assign inputs to distinct 2D and 3D sub-tasks. Similarly, HandOver~\cite{tutuncu2025handover} addressed transition fluidity by unifying mouse precision with hand-tracking expressiveness for seamless target selection and manipulation. These findings underscore the need to extend the mouse's interaction space into a 3D environment to minimize unnecessary device transitions. 

Human spatial manipulation is fundamentally guided by visual attention~\cite{pfeuffer2017gaze}. Users naturally look at an object before selecting or interacting with it. Grounded in this natural eye-hand coordination process, we incorporate a gaze cursor into our approach. Furthermore, this design seamlessly aligns with the current trajectory of modern XR headsets (e.g., Vision Pro), which are increasingly adopting gaze-based interactions and multimodal techniques such as Gaze+Pinch~\cite{kim2025pinchcatcher, pfeuffer2024design}. Moreover, we focused on 3D translation, a fundamental task in spatial manipulation~\cite{wagner2025Pen, tutuncu2025handover}, performed via a mouse. Overall, we introduce \papername, a technique that dynamically modulates translation mapping based on gaze and mouse cursor alignment (Fig~\ref{fig:teaser}), inspired by Gaze-Shifting~\cite{pfeuffer2015gaze}. In this approach, the spatial overlap between the gaze and the mouse cursor serves as a trigger to switch between vertical (XY) and horizontal (XZ) translation.

While the mouse and keyboard remain the standard for efficient 2D interaction, adapting these familiar metaphors to gaze-based systems poses challenges due to inherent gaze imprecision and the need to align with users' mental models. Therefore, our primary objective is to systematically explore the foundational design space of gaze-mouse coordination and optimize these variables to design a seamless and continuous 3D translation without modality switching. Specifically, we focus on translation mapping profiles (overlapXZ vs. overlapXY) and gaze cursor aperture (small vs. large). Moreover, we evaluated \papername against a standard 3D Gizmo, which serves as the ubiquitous baseline for spatial translation in both 2D desktop and XR applications~\cite{tutuncu2026world, gustavsson2014interaction}, to validate its practical efficacy.

The results demonstrated a trade-off regarding the gaze cursor aperture. Increasing the cursor size significantly reduced the required hand movement, but it also introduced ambiguity that led to increased selection errors. Also, the \papername technique significantly improved interaction efficiency and user satisfaction compared to the standard Gizmo, while maintaining comparable placement precision. Subjective evaluations further supported these findings, indicating that \papername induced lower cognitive load and achieved higher user satisfaction and usability scores than the Gizmo. Notably, for tasks requiring fine depth translation, the overlapXZ mapping, which activates the depth plane upon cursor overlap, was preferred predominantly over the overlapXY mapping. 

Based on these findings, this study offers two primary contributions. First, we systematically explore and quantify the foundational design space of gaze-mouse coordination and verify that the overlapXZ mapping strongly aligns with users' natural mental models of spatial translation. Second, we empirically validate that \papername, in its optimal configuration, improves interaction efficiency and reduces cognitive load compared to the standard 3D Gizmo.

%%%%%%%%%%%%%%%%%%%%%%%%%%%%%%%%%%%%%%%%%%%%%%%%%%%%%%%%%%%%%%%%%%%%%%%%%%%%%%%%%%%%%%%%%%%%%%%%%%

\section{Related Work}
Given the decades-long dominance of the keyboard and mouse as the standard for efficiency and familiarity in computing, adapting their usage to 3D environments has been a persistent research focus \cite{balakrishnan1997rockin, hubenschmid2025spatialmouse}. However, a fundamental challenge lies in the dimensionality mismatch between the 2D planar input of a mouse and the 3D spatial output required in VR or AR. Mendes et al.~\cite{mendes2019survey} note that to bridge this gap, 3D manipulation techniques frequently employ task decomposition, reducing complex 6-degree-of-freedom (DoF) transformations into lower-dimensional operations. Implementation of such decomposition is often realized through constraint-based mechanics. While standard widgets such as Gizmos typically constrain movement to a single axis~\cite{tutuncu2026world}, the core principle remains that reducing the number of simultaneously active DoF enhances control. Drey et al.~\cite{drey2023investigating} further support the efficacy of this constraint-based interaction; their findings indicate that restricting simultaneous DoF, whether to a single axis or a specific plane, allows users to perform tasks with greater precision by significantly reducing the cognitive load associated with full multi-axis control.

Beyond theoretical precision, empirical studies in cross-reality contexts further validate the practical advantages of the mouse~\cite{zhou2022depth, grubert2020back}. Cools et al.~\cite{cools2025comparison} compared four techniques (i.e., Mouse, Hand, Modality Switch (mandating a change in input device), and Modality Choice (allowing voluntary input selection)) in a cross-reality based task, transferring and adjusting 3D objects between XR and 2D screen space. They reported that the Mouse-based technique achieved the fastest performance due to users' familiarity with the device, while Modality Choice was the most preferred, as it enabled users to strategically assign the most suitable input modality to 2D and 3D sub-tasks, respectively. However, relying on voluntary switching still imposes cognitive load and physical effort, suggesting a need for a unified input method that minimizes frequent transitions~\cite{surale2019experimental, wentzel2024switchspace}.

To fulfill the need for a unified interface, researchers have explored methods to directly integrate the mouse into VR/AR environments~\cite{grubert2020back, hubenschmid2025spatialmouse}. HandOver~\cite{tutuncu2025handover} unifies mouse precision with hand-tracking expressiveness; it enables users to select targets with a depth-aware mouse cursor and seamlessly switch to direct 3D manipulation by simply hovering their hand over the device. Their study showed that this technique yielded lower task errors across all distances compared to traditional ray-casting and improved interaction ergonomics by allowing users to rest their hands during the selection phase. Taking a device-centric approach, Hubenschmid et al.~\cite{hubenschmid2025spatialmouse} introduced SpatialMouse, a hybrid device that functions as a standard optical mouse on a surface but seamlessly transitions into a spatial controller when lifted. This design eliminates the need for device switching, enabling users to fluidly move between precise 2D pointing and 3D spatial manipulation within a single workflow. However, these hybrid solutions still require physical transitions, either lifting the hand or the device itself, which can disrupt the workflow continuity compared to a purely hands-on-mouse approach.

%%%%%%%%%%%%%%%%%%%%%%%%%%%%%%%%%%%%%%%%%%%%%%%%%%%%%%%%%%%%%%%%%%%%%%%%%%%%%%%%%%%%%%%%%%%%%%%%%%

\begin{figure*}
  \centering
  \includegraphics[width=0.85\textwidth]{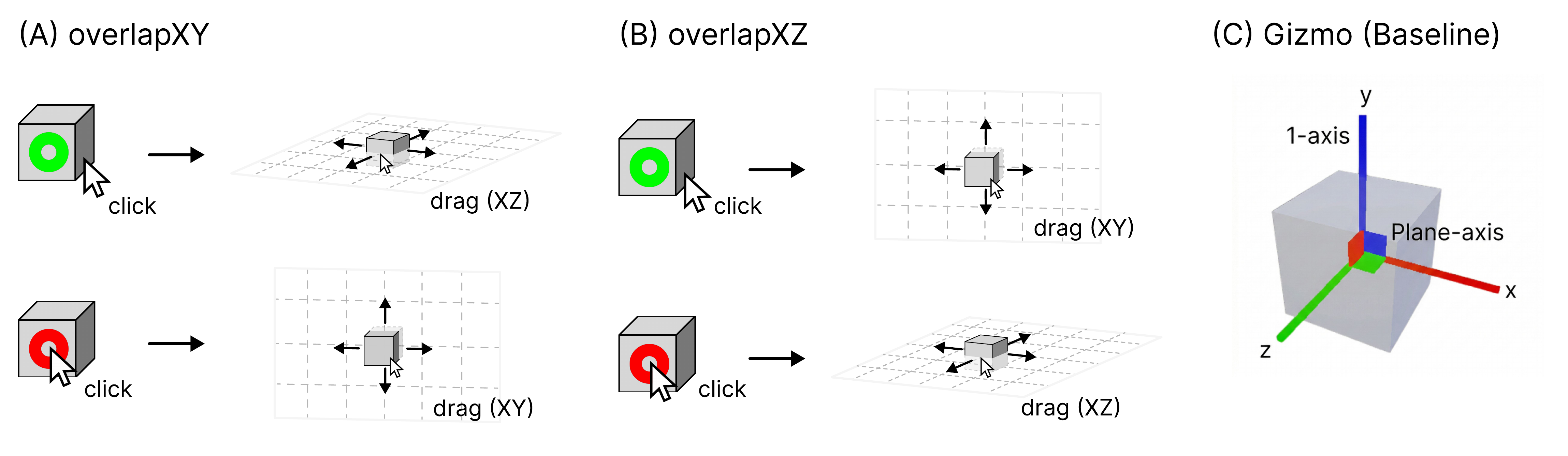}
  \caption{Overview of \papername (overlapXY and overlapXZ) and baseline 3D Gizmos translation technique. (a) overlapXY: When gaze and mouse cursors are disjoint, the object translates on the XZ (depth) plane; overlapping cursors activate the XY plane as typical mouse movement in 2D space. (b) overlapXZ: Employing an inverted mapping, this mode triggers XZ plane translation upon gaze-cursor overlap. (c) Gizmo implementation: Bar handles control axis translation, while square handles enable planar manipulation.}
  \label{fig:methods}
\end{figure*}

\section{Design of Spatial Translation via Gaze-Mouse Cursor Alignment}
\subsection{Design Considerations}
\subsubsection{Extending Mouse Input for 3D Translation}
To design an interaction technique that eliminates the need for repetitive physical transitions between input methods, we first reviewed previous work to gain insight. Previous studies have investigated techniques for transferring objects across heterogeneous environments, such as translating items across screen boundaries or switching between input modalities like mouse and hand tracking, depending on the workspace~\cite{rau2025traversing, cools2025comparison}. While effective in their respective domains, these methods inherently treat the desktop and XR environments as distinct zones that require explicit bridging actions. These were more focused on augmenting 2D content in a 3D space by incorporating an additional step that allows users to perform the transition. In addition, each input technique remained constrained to its specific interface. Consequently, users were frequently compelled to switch between devices, utilizing the mouse for precise 2D tasks and hands or controllers for 3D manipulation. This reliance on multiple devices fragments the workflow and increases physical effort. Thus, we aimed to design an interaction technique that extends the interaction area of the mouse into 3D space. This approach could allow users to manipulate objects within the 3D environment using only the mouse, thereby eliminating the need for physical transition between input devices.

\subsubsection{Facilitating Mental Models for Spatial Interaction}
Effective interface design relies on minimizing the Gulf of Execution between a user's internal goals and the system~\cite{norman1986cognitive}. To bridge this gap, the spatial mapping between input controls and display responses must align with users' cognitive expectations, adhering to the principle of stimulus-response compatibility~\cite{fitts1953sr, brebner1972spatial}. Establishing an accurate mental model is particularly challenging in heterogeneous 2D-3D environments due to dimensional differences~\cite{li2023effect}. Thus, interactions must evolve beyond simple technical mappings to reflect the user's natural spatial intentions~\cite{cheng2025mental}. Building on these theoretical foundations, we propose an interaction technique that effectively bridges the Gulf of Execution by ensuring 2D mouse inputs correspond intuitively to intended 3D manipulations.

\subsection{\papername Interaction Design}
Based on these considerations, we propose \papername, a technique inspired by the Gaze-Shifting mechanism~\cite{pfeuffer2015gaze} that modulates the translation plane based on the alignment of gaze and mouse input. As illustrated in Fig~\ref{fig:teaser} and~\ref{fig:methods}, the mouse cursor serves as the primary tool for direct manipulation, while the gaze cursor functions as an auxiliary trigger for mode switching. Specifically, the system detects whether the two cursors coincide in space to lock the corresponding coordinate axis, allowing the user to manipulate the object within the defined plane. Within this framework, we designed two mapping conditions (Fig~\ref{fig:methods}): \textit{overlapXY} and \textit{overlapXZ}. In the \textit{overlapXY} condition, the system activates the vertical XY plane when the cursors overlap and defaults to the horizontal XZ plane when they are disjoint. The \textit{overlapXZ} condition employs the inverse logic.

Furthermore, we integrated visual feedback mechanisms into the system to let users be aware of the current state. Specifically, to address the inherent challenges of perceiving object movement in 3D environments~\cite{mendes2019survey}, we implemented a grid plane visualization similar to the approach employed by Wagner et al.~\cite{wagner2024eye}. Without explicit spatial cues, users often misjudge the depth and relative size between the target and the destination, leading to perspective illusions where separated objects appear aligned~\cite{mendes2019survey}. To mitigate this, our design features a semi-transparent plane intersecting the target's center. This visual guide not only clarifies the object's movement trajectory but also enhances depth perception, enabling users to accurately verify spatial overlap and employ precise target adjustment strategies.

In addition to spatial cues, we provided explicit feedback regarding the cursor's behavior to clarify the interaction mode. To minimize visual distraction and allow the user to focus on manipulation, the gaze cursor is rendered invisible while the object is being selected and dragged, reappearing immediately upon the completion of the operation. Simultaneously, the system indicates the overlap status through color-coded cues. As illustrated in Fig~\ref{fig:teaser}, the gaze cursor changes from gray (default) to green upon target acquisition and turns red to signify a valid overlap. The combination of visibility control and color changes facilitates maintaining a clear understanding of the current interaction status throughout the task.

Lastly, in terms of gaze interaction, we applied a gaze snapping and locking technique to minimize selection errors caused by eye tracking noise, thereby allowing the study to focus solely on the efficacy of the proposed gaze-mouse coordination strategy~\cite{yu2025deriving, kim2025pinchcatcher, lystbaek2024hands}. The gaze cursor locks to the center of the target object upon a fixation duration ($T_{dwell}$) of 0.5s and disengages when the gaze exits the target's boundaries. Specifically, the system calculates the angular difference $\theta$ between the vector extending from the gaze origin $O$ to the target center $P_t$ and the current gaze direction vector $\vec{d}_{gaze}$ as shown in Equation~\ref{eq:angle_diff}:

\begin{equation}
\label{eq:angle_diff}
\theta = \arccos \left( \frac{\vec{d}_{gaze} \cdot (P_t - O)}{ \| \vec{d}_{gaze} \| \| P_t - O \| } \right)
\end{equation}

If the angle $\theta$ falls within the activation threshold ($\theta_{locked} = 5^{\circ}$), which is derived from the representative target size, the cursor automatically locks to the center of the target. Upon locking, the target remains active based on a hysteresis mechanism to maintain stability against eye jitter or tracking errors~\cite{velloso2017motion}. The locking state $F_{lock}(t)$ at time $t$ is determined by Equation~\ref{eq:hysteresis}, where 1 represents the locked state and 0 the released state. This mechanism employs asymmetric thresholds with the release threshold $\theta_{release}$ set to $10^{\circ}$, corresponding to twice the target size:

\begin{equation}
    F_{lock}(t) = 
    \begin{cases} 
      1 & \text{if } \theta \leq \theta_{locked} \text{ for } \Delta t \geq T_{dwell} \\
      0 & \text{if } \theta > \theta_{release} \\
      S_{lock}(t-1) & \text{otherwise}
    \end{cases}
    \label{eq:hysteresis}
\end{equation}

This mechanism could facilitate conveying a clear intent to manipulate the specific object while maintaining interaction stability, even if the gaze slightly drifts.

%%%%%%%%%%%%%%%%%%%%%%%%%%%%%%%%%%%%%%%%%%%%%%%%%%%%%%%%%%%%%%%%%%%%%%%%%%%%%%%%%%%%%%%%%%%%%%%%%%

\section{User Study Methods}
To identify the optimal interaction configuration, we evaluated the effects of two independent variables in a user study: gaze-cursor aperture and translation mapping profile. Addressing the inherent precision limitations of current HMD-embedded eye trackers~\cite{hou2024unveiling}, two aperture sizes were tested: $1.65^\circ$ (Small) and $3.31^\circ$ (Large). The small aperture corresponded to approximately $1/3$ of the target diameter, while the large aperture was double that size. Regarding mapping profiles, we compared two mode-switching strategies: \textit{overlapXY} and \textit{overlapXZ}. Exploring this design space was critical to determine which mapping best aligns with users' mental models established by standard 2D interaction metaphors.

While recent cross-reality interaction techniques~\cite{tutuncu2025handover, rau2025traversing} have introduced novel methods for transitioning between 2D and 3D, they primarily focus on target selection or employ hybrid modalities that require users to physically release the mouse to perform mid-air hand gestures. Directly comparing a \papername against these hybrid methods within an isolated spatial translation task could introduce an asymmetric comparison. Moreover, the current paper focuses on the context of keeping the user's hand continuously on the mouse. Therefore, we employed a standard 3D Gizmo (Fig~\ref{fig:methods} (C)) as our baseline. This ubiquitous paradigm has been widely validated across both traditional 2D platforms and XR environments~\cite{drey2023investigating, tutuncu2026world, gustavsson2014interaction}, enabling explicit constrained movement via axis and planar handles. %Notably, even in recent studies where the mouse is utilized as a cross-reality cursor beyond 2D screen boundaries (e.g., World Mouse~\cite{tutuncu2026world}), Gizmos remain the standard approach for executing 3D spatial translation.

\subsection{Task}
Grounded in prior work~\cite{tutuncu2025handover}, our study employed a 3D translation task where users drag a cube from a generation point to a target destination. This design aims to replicate cross-device scenarios in which objects are migrated from a 2D display to the immediate 3D environment~\cite{cools2025comparison}. Such interactions resemble practical workflows, including dragging 3D models beyond screen boundaries~\cite{zhou2022depth} or arranging widgets (e.g., music player, memo app) along the monitor's periphery. To visually ground this context, we incorporated a gray 4:3 screen at the center to represent a physical monitor, as shown in Fig~\ref{fig:task}.

The study also manipulated the translation amplitude by employing two distance levels ($30^\circ$ and $40^\circ$ radius). Targets were positioned along a circular trajectory centered $1\,\text{m}$ anterior to the user's head position. Furthermore, depth variability was introduced via random offsets of $\pm0.1\,\text{m}$ for the $30^\circ$ amplitude and $\pm0.2\,\text{m}$ for $40^\circ$ amplitude, thereby enforcing volumetric manipulation beyond planar movements. Each target was rendered as an $8.73\,\text{cm}$ cube and presented sequentially.

\subsection{Procedure}
\begin{figure}
  \includegraphics[width=0.95\linewidth]{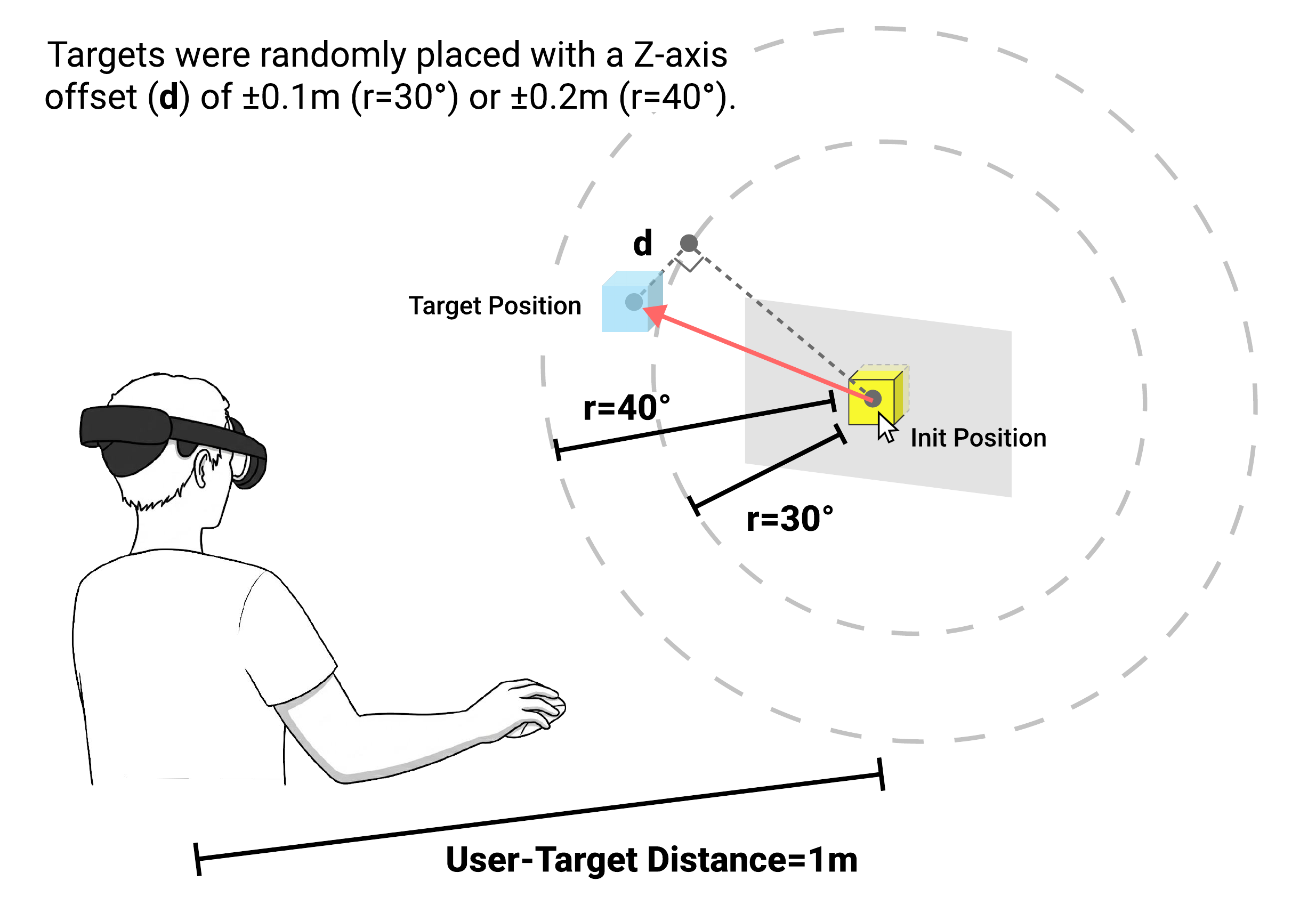}
  \centering
  \caption{Illustration of the task environment. The targets and screen are positioned 1 m from the user. Targets are arranged in a circular pattern with a radius ($r$) of $30^\circ$ or $40^\circ$, featuring a z-axis depth offset ($d$). The depth offset is set randomly to $\pm0.1$m for the $30^\circ$ radius and $\pm0.2$m for the $40^\circ$ radius. The task requires users to drag the yellow cube object to the blue target destination.}
  \label{fig:task}
\end{figure}

Upon arrival, participants provided informed consent and completed a demographic questionnaire, and eye-tracking calibration was performed on the HMD before the experiment. Then, a five-minute training session was administered to familiarize participants with the interaction techniques. In the main experiment, the mouse cursor was re-centered at the beginning of each trial. Participants were instructed to translate a cube object from the center screen to a target location as quickly and accurately as possible using the allocated technique (Fig~\ref{fig:task}). To confirm the placement, participants clicked the right mouse button. Visual feedback was provided to assist accuracy; the target turned green when the cube object overlapped the target volume by more than 85\%. Following confirmation, participants proceeded to the next trial at their own pace by clicking the right mouse button again.

We employed a within-subjects design with the order of the five techniques (2 gaze-cursor apertures $\times$ 2 translation mapping profiles, baseline) fully randomized using the Fisher-Yates Shuffle~\cite{schafer2021controlling}. Each technique block consisted of 20 repetitions of the trials with two translation amplitudes, resulting in a total of 200 trials per participant (5 techniques $\times$ 2 amplitudes $\times$ 20 repetitions). Subjective feedback was collected via questionnaires after each technique block. The entire session lasted approximately 90 minutes.

\subsection{Experimental Setup and Implementation}
The study was implemented using Unity (2022.3.57f1) and the Meta XR-all-in-One SDK (v72) on a Meta Quest Pro (90 Hz, 111.24° FoV), equipped with its embedded eye tracker (30 Hz). We applied a 1\texteuro~filter on gaze and hand movement to reduce noise and smooth the tracking input~\cite{casiez20121, Zhang22}. The parameters were set to $f_{c_{min}}$=0.9 and $\beta$=15 for gaze and $f_{c_{min}}$=0.9 and $\beta$=90 for hand movement, as in the previous Gaze-based interaction study~\cite{wagner2024eye}. Mouse movement acceleration was calculated based on 1000 DPI by multiplying the value obtained using Unity's \textit{GetAxis} function by a sensitivity coefficient of 20 and accumulating it on a per-frame basis. We aimed to simulate the realistic and natural context of mouse movement on the virtual canvas.

\subsection{Evaluation Metrics}

    \noindent\textbf{Task Completion Time (TCT)}: The elapsed time from the initialization of a trial to the final confirmation. 

    \noindent\textbf{Target Position Offset}: The spatial precision of placement, measured as the Euclidean distance between the target destination's center and the object's final position upon task completion. A lower offset value signifies greater precision in alignment.

    \noindent\textbf{Mouse Click Count}: The total number of mouse click events registered during the object manipulation process. A lower click count indicates fewer discrete steps or corrective actions required to complete the task.
    
    \noindent\textbf{Click Error Rate}: The precision of the interaction was evaluated by the proportion of failed selection attempts. This is defined as the ratio of erroneous clicks (i.e., clicking on empty space instead of the target object or handle) to the total number of selection attempts per trial.

    \noindent\textbf{Hand Movement}: The cumulative Euclidean distance of the user's right-hand trajectory required to complete the task. The movement was recorded using the hand tracking feature of the Meta SDK.

    \noindent\textbf{Questionnaires}: We employed standard questionnaires to evaluate subjective user experience through the System Usability Scale (SUS)~\cite{brooke2013sus} and the NASA-TLX~\cite{hart2006nasa}. Additionally, participants reported their satisfaction level~\cite{yu2020fully}. All questionnaires were collected using a 7-point Likert scale.

    \noindent\textbf{Ranking}: Upon completion of the experiment, participants ranked the techniques based on their preference.

\subsection{Participants}
We recruited a total of 20 participants (12 Female, 8 Male, Age M=26.05, SD=2.58) from a local university. Among them, four participants wore glasses, and 19 identified as right-handed. All participants stated that they had prior experience with VR. Specifically, 17 reported experience with VR games, and 10 had used VR more than 10 times. All experimental procedures and methods were approved by the university's Institutional Review Board (IRB), and participants were rewarded \$10.

%%%%%%%%%%%%%%%%%%%%%%%%%%%%%%%%%%%%%%%%%%%%%%%%%%%%%%%%%%%%%%%%%%%%%%%%%%%%%%%%%%%%%%%%%%%%%%%%%%

\begin{figure*}
  \includegraphics[width=0.95\textwidth]{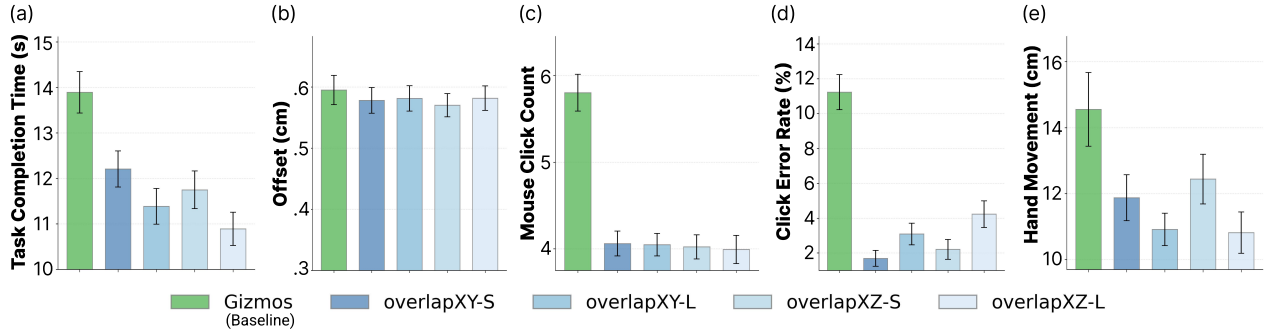}
  \centering
  \caption{Averaged performance metrics for each interaction technique: (a) Mean Task Completion Time, (b) Mean Target Position Offset, (c) Mean Mouse Click Count, (d) Mean Click Error Rate, (e) Mean Hand Movement Distance. Error bars represent 95\% confidence intervals.}
  \label{fig:result}
\end{figure*}

\section{Results}
Of the 4,000 trials collected, 3,876 were retained for analysis, with 124 (3.1\%) excluded for failing to meet the target threshold. Task failures were defined as instances where the distance between the object and target centers exceeded the target radius. Additionally, data were aggregated across translation amplitudes to isolate the general effects of the interaction techniques.

We first compared the baseline (Gizmos) with the four \papername interaction types using paired T-tests. Variables that violated the normality assumption, such as Hand Movement, were analyzed using the nonparametric Wilcoxon signed-rank test. To examine the effects within \papername interactions, we conducted a two-way repeated measures ANOVA (Mapping Profile $\times$ Gaze Cursor Size) on data transformed using the Aligned Rank Transform (ART)~\cite{wobbrock2011aligned}. Subjective ratings (e.g., NASA-TLX, SUS) were evaluated using the nonparametric Friedman test followed by Conover’s post-hoc tests. Significance levels for all post-hoc pairwise comparisons were adjusted using the Bonferroni correction.

\subsection{Task Completion Time (Fig~\ref{fig:result} (a))}
The baseline Gizmos condition ($M=13.894, SD=6.429$) resulted in significantly longer task completion times compared to all \papername techniques. The most substantial time reduction was observed in overlapXZ-large ($M=10.981, SD=5.206$), which yielded the largest deviation from the baseline ($\Delta=3.003, t(19)=4.311, p=.002$). Significant improvements were also found in overlapXY-large ($M=11.386, SD=5.574$; $\Delta=2.508, t(19)=3.299, p=.015$) and overlapXZ-small ($M=11.749, SD=5.901$; $\Delta=2.145, t(19)=3.591, p=.008$). overlapXY-small ($M=12.206, SD=5.668$) showed the most modest improvement, with the smallest but still significant difference from the Gizmos ($\Delta=1.688, t(19)=2.779, p=.048$).

In the comparison among \papername design factors, no significant main effects were observed for Mapping Profile ($F_c(1, 57)=1.457, p=.232, \eta_p^2=.025$) or Gaze Cursor Size ($F_c(1, 57)=2.569, p=.115, \eta_p^2=.043$). Additionally, no significant interaction effect was observed among the factors ($F_c(1, 57)=.201, p=.656, \eta_p^2=.004$).

\subsection{Target Position Offset (Fig~\ref{fig:result} (b))}
Analysis of target position offset revealed no significant differences between the Gizmo condition ($M=0.595, SD=0.336$) and any of the proposed techniques ($t(19)s<1.390, ps>.05$). Specifically, for the small cursor variants, overlapXZ-small ($M=0.570, SD=0.271$) and overlapXY-small ($M=0.578, SD=0.299$) exhibited comparable precision to the baseline. Likewise, the large cursor variants maintained similar precision levels, with overlapXZ-large ($M=0.582, SD=0.283$) and overlapXY-large ($M=0.581, SD=0.295$) showing no statistical deviation.

In the comparison among combinations, the results showed no main effects for Mapping Profile ($F_c(1, 57)=0.210, p=.648, \eta_p^2=.004$) or Gaze Cursor Size ($F_c(1, 57)=0.345, p=.560, \eta_p^2=.006$). Furthermore, no interaction effect was observed ($F_c(1, 57)=0.001, p=.978, \eta_p^2<.001$). These results demonstrate that \papername maintains baseline-level placement precision, remaining consistent regardless of the mapping profile or cursor size.

\subsection{Mouse Click Count (Fig~\ref{fig:result} (c))}
Analysis of click frequency revealed that the Gizmos condition ($M=5.803,\ SD=2.994$) incurred significantly more attempts compared to all proposed techniques ($ps<.001$). Specifically, the largest reduction in click count was observed in overlapXZ-large ($M=3.991,\ SD=2.312$; $\Delta=1.833,\ t(19)=7.514$), followed by overlapXZ-small ($M=4.022,\ SD=1.977$; $\Delta=1.796,\ t(19)=7.719$). The overlapXY variants showed similar improvements, where overlapXY-large ($M=4.048,\ SD=1.844$) and overlapXY-small ($M=4.060,\ SD=2.042$) yielded reductions of $1.777$ ($t(19)=6.669$) and $1.767$ ($t(19)=6.930$) relative to the baseline, respectively.

In the comparison among combinations, the results showed no main effects for Mapping Profile ($F_c(1, 57)=.370,\ p=.545,\ \eta_p^2=.006$) or Gaze Cursor Size ($F_c(1, 57)=.045,\ p=.833,\ \eta_p^2<.001$). Furthermore, no interaction effect was observed ($F_c(1, 57)=.088,\ p=.768,\ \eta_p^2=.002$).

\subsection{Click Error Rate (Fig~\ref{fig:result} (d))}
The Gizmos condition ($M=11.232,\ SD=14.112$) resulted in a significantly higher rate of incorrect clicks in empty space compared to all proposed techniques ($t(19)s>5.46,\ ps<.001$). Specifically, the most substantial reduction in error rate was observed in overlapXY-small ($M=1.686,\ SD=6.516; \Delta=9.546$), followed by overlapXZ-small ($M=2.206,\ SD=8.103; \Delta=9.026$). The Large variants also showed significant accuracy improvements: overlapXY-large ($M=3.092,\ SD=8.834; \Delta=8.140$) and overlapXZ-large ($M=4.231,\ SD=10.872; \Delta=7.001$).

Regarding the comparison among \papername design factors, we found a significant main effect of Mapping Profile on click error rate ($F_c(1, 57)=5.907,\ p=.018,\ \eta_p^2=.094$). Post-hoc analysis revealed that the error rate for overlapXZ ($M=3.219,\ SD=9.639$) was significantly higher than that of overlapXY ($M=2.387,\ SD=7.788$) ($t(57)=2.431,\ p=.018$). We also observed a significant main effect of Gaze Cursor Size ($F_c(1, 57)=20.474,\ p<.001,\ \eta_p^2=.264$). Results indicated that the error rate was significantly higher with the Large cursor ($M=3.661,\ SD=9.918$) than with the Small cursor ($M=1.944,\ SD=7.351$) ($t(57)=4.525,\ p<.001$), suggesting that target selection was more accurate with the smaller cursor. No significant interaction effect was found between the two factors ($F_c(1, 57)=.659,\ p=.420,\ \eta_p^2=.011$).

\subsection{Hand Movement (Fig~\ref{fig:result} (e))}
Data from P05 were excluded from the analysis only due to tracking failures affecting 32 trials (16\%). Analysis revealed that the Gizmos condition ($M=14.583,\ SD=5.287$) required significantly more hand movement compared to most proposed techniques. Specifically, significant reductions were observed in overlapXZ-large ($M=10.853,\ SD=3.613; Z=3.240,\ p=.001,\ \Delta=3.730$), overlapXY-large ($M=10.920,\ SD=3.427; Z=2.958,\ p=.003,\ \Delta=3.663$), and overlapXY-small ($M=11.869,\ SD=4.215; Z=2.837,\ p=.005,\ \Delta=2.714$). However, the reduction in overlapXZ-small ($M=12.547,\ SD=5.336,\ \Delta=2.036$) did not reach statistical significance ($Z=1.912,\ p=.056$).

In the comparison among combinations, the results showed no significant main effects for Mapping Profile ($F_c(1, 54)=.006,\ p=.937,\ \eta_p^2<.001$). The main effect of Gaze Cursor Size approached significance but did not meet the threshold ($F_c(1, 54)=3.916,\ p=.053,\ \eta_p^2=.068$). Furthermore, no interaction effect was observed ($F_c(1, 54)=.042,\ p=.838,\ \eta_p^2<.001$).

\subsection{Questionnaires (Fig~\ref{fig:questionnaire})}
Analysis of NASA-TLX showed a main effect of technique on all subscales, except for Temporal Demand. Post-hoc comparisons showed that the Gizmos condition required significantly more Effort ($\chi^2(4)=16.90, p=.002$) and Mental Demand ($\chi^2(4)=19.86, p<.001$) than the overlapXZ variants and overlapXY-small ($t(76)s>3.065, ps<.030$). Furthermore, the Gizmos induced significantly higher Frustration ($\chi^2(4)=21.97, p<.001$) and Physical Demand ($\chi^2(4)=25.85, p<.001$) compared to all proposed techniques ($t(76)s>3.760, ps<.003$). The Satisfaction ($\chi^2(4)=31.41, p<.001$) score was significantly lower for the Gizmos than for all other conditions ($t(76)s>4.225, ps<.001$). Regarding Performance ($\chi^2(4)=13.52, p=.009$), participants rated the overlapXZ conditions (both Small and Large) significantly higher than the Gizmos ($t(76)s=3.265, ps=.016$).

The analysis of perceived usability scores revealed a significant main effect of technique ($\chi^2(4)=16.22, p=.003$). Post-hoc comparisons indicated that the Gizmos (M=3.125, SD=1.010) condition scored significantly lower than both overlapXZ-small (M=4.183, SD=0.960; $t(76)=3.402, p=.011$) and overlapXZ-large (M=4.200, SD=4.098; $t(76)=3.960, p=.002$). Notably, the scores for all five techniques (overlapXY-small M=3.775, SD=0.972; overlapXY-large M=3.833, SD=1.067), including the Gizmos, exceeded 3.10, meeting the minimum threshold for acceptable usability~\cite{sauro2016quantifying}. Among them, the overlapXZ variants achieved scores above 4.08, surpassing the average usability benchmark~\cite{sauro2016quantifying}.

\subsection{Ranking}
\begin{figure*}
  \includegraphics[width=\textwidth]{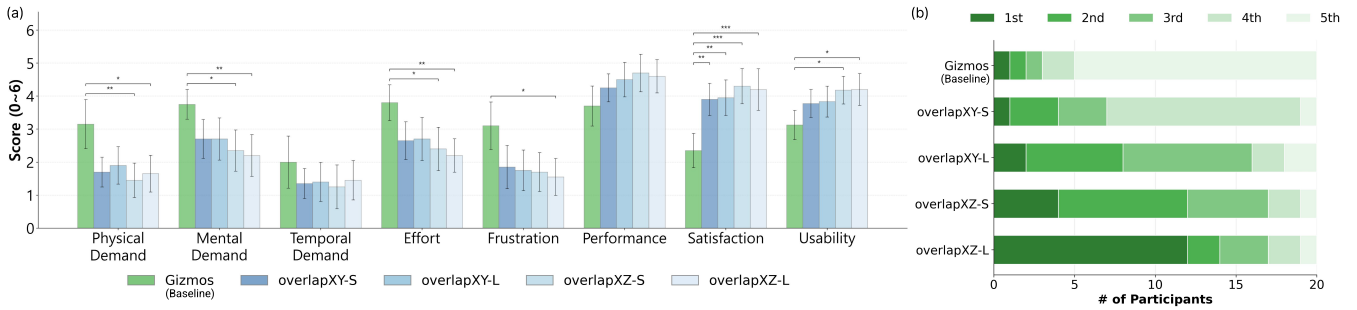}
  \caption{(a) NASA-TLX workload, perceived satisfaction, and usability ratings on a 7-point Likert scale (0-6) for each technique. The error bars represent 95\% confidence intervals. Statistical significance is represented by * for $p < .05$, ** for $p < .01$, and *** for $p < .001$. (b) The result of the participants’ ranking for each technique.}
  \label{fig:questionnaire}
\end{figure*}

Among the five techniques, overlapXZ-large emerged as the most preferred, with 12 participants ranking it first (Fig~\ref{fig:questionnaire}). In contrast, Gizmos was the least preferred technique, with 15 participants ranking it fifth. Moreover, overlapXZ-small was frequently ranked higher than overlapXY-large, while overlapXY-small was most frequently ranked fourth. These results indicate that the overlapXZ mode was the preferred mapping profile mechanism and that the larger gaze cursor was preferred over the smaller one. 

%%%%%%%%%%%%%%%%%%%%%%%%%%%%%%%%%%%%%%%%%%%%%%%%%%%%%%%%%%%%%%%%%%%%%%%%%%%%%%%%%%%%%%%%%%%%%%%%%%

\section{Discussion}
\subsection{Optimization of \papername Design}
\subsubsection{Mental Model Alignment in Spatial Mapping}
Through our design space exploration, a key finding of our study is the strong user preference for the overlapXZ mapping profile. One possible explanation for this preference is its potential alignment with users' mental models of intent and action~\cite{cheng2025mental}, which resonates with the principle of stimulus-response compatibility~\cite{fitts1953sr, brebner1972spatial}. Participants perceived the non-overlapping state as the default for planar (XY) arrangement, which aligns with the dominance of 2D tasks in cross-reality workflows. Conversely, the act of aligning the mouse with the gaze was conceptually linked to a heightened state of intent, a deliberate activation step distinct from the passive nature of planar manipulation. This suggests that matching a higher-effort input (deliberate alignment) with an auxiliary output (depth translation) may create a natural congruence, helping to bridge the gulf between user intent and system response.

In contrast, the overlapXY condition seemed to conflict with habitual motor strategies. Since planar mouse movements are typically performed subconsciously, requiring conscious effort (gaze-mouse alignment) for these fundamental tasks was often found counterintuitive. This mismatch may impose unnecessary cognitive friction. Consequently, our findings could imply that intentional triggers in cross-dimensional interaction could be better mapped to auxiliary actions (e.g., depth translation) to support stimulus-response compatibility.

\subsubsection{Trade-off in Gaze Cursor Aperture}
Our investigation into gaze cursor size revealed a distinct trade-off between precision and ease of use. The small gaze cursor resulted in lower selection error rates, confirming that tighter tolerance filters out ambiguous inputs. However, this incurred the cost of increased physical effort, as evidenced by longer hand movement trajectories. In contrast, the large cursor significantly reduced the required motor space. For instance, the overlapXZ-large condition required only 10.85 cm of hand movement, which is a 3.73 cm reduction from the baseline. By reducing the effort for precise hand-eye coordination, the larger aperture led to higher subjective satisfaction and comfort scores. Conversely, the large cursor also introduced the Midas Touch problem~\cite{penkar2012designing}, where unintended overlaps triggered accidental depth activations. Despite the higher error rate, the majority of participants preferred the larger aperture, prioritizing fluid interaction over strict error prevention. This observation suggests that in translation tasks where corrective actions are relatively low-cost, users favor reduced motor demand.

Prior literature on gaze-based interaction has largely circumvented the issue of tracking noise by simply expanding target colliders~\cite{kim2025pinchcatcher, lystbaek2024hands}, leaving the specific design space of gaze cursor aperture relatively unexplored. Consequently, our study addresses this gap by providing empirical evidence that the optimal cursor size is not a fixed parameter but is highly dependent on the functional context. Specifically, there is a trade-off between minimizing physical effort to facilitate rapid translation activation and maximizing input fidelity to ensure precise cursor positioning. This insight could serve as a foundational guideline for future improvements, suggesting that it should move towards context-aware adaptive mechanisms that dynamically modulate cursor tolerance to reconcile the conflicting demands of stability and maneuverability.

\subsection{Efficiency of Gaze-Mouse Coordination}
By applying the optimal configuration identified in our design space exploration, \papername yielded substantial improvements over the standard 3D Gizmo across all evaluated metrics. \papername yielded a substantial reduction in task completion time (up to a 3s improvement for overlapXZ-large) and required significantly fewer mouse clicks compared to the Gizmo. This performance gain can be attributed to the elimination of the explicit mode-switching overhead inherent in traditional 3D Gizmos.

Standard Gizmos rely on constraint-based decomposition, requiring users to explicitly select specific axes or planar handles to constrain movement. This introduces a secondary task that interrupts the primary workflow of object manipulation. Furthermore, we found that the act of selecting these handles is often hindered by perspective distortion in 3D space; variations in viewing angle and object depth can make the gizmo difficult to acquire accurately. In contrast, \papername utilizes gaze as a natural, implicit modality to modulate the translation plane dynamically. By mapping the spatial overlap of the gaze and mouse cursors to the depth axis, our technique effectively extends the mouse’s input space without requiring physical device transitions or explicit UI toggle operations for mode change.

Subjective evaluations further support these quantitative findings. The NASA-TLX results indicated that the Gizmos condition induced significantly higher physical and mental demand compared to \papername. By mitigating these constraints through a larger, gaze-responsive activation area, \papername successfully bridges the Gulf of Execution~\cite{norman1986cognitive}, allowing users to translate 2D input into 3D intent with reduced cognitive load.

%%%%%%%%%%%%%%%%%%%%%%%%%%%%%%%%%%%%%%%%%%%%%%%%%%%%%%%%%%%%%%%%%%%%%%%%%%%%%%%%%%%%%%%%%%%%%%%%%%

\section{Limitations \& Future work}
The current evaluation focused on exploring the foundational design space of gaze-mouse coordination specifically for 3D translation. Since our primary scope was intentionally limited to continuous spatial translation without device switching, we evaluated \papername against a standard 3D Gizmo to isolate the performance of mapping profiles and cursor aperture. However, this baseline comparison unavoidably conflates the effects of gaze coordination with differences in effective target size, as replacing thin Gizmo handles with a large object selection area and generous gaze-snapping tolerance may have facilitated target acquisition. We also utilized a single fixed-size object from a stationary viewpoint; manipulating significantly smaller targets could make maintaining the required $0.5\,\text{s}$ fixation challenging, while restricting users from panning likely exacerbated the perspective distortion experienced during Gizmo manipulation. 

Next, we did not conduct empirical comparisons with hybrid interaction techniques~\cite{tutuncu2025handover, rau2025traversing} or a size-matched, gaze-independent baseline. Since hybrid techniques and \papername likely possess distinct contextual strengths depending on the specific task requirements (e.g., targeting vs. continuous moving, or the tolerable cost of device switching), future work must investigate these techniques within complex, high-frequency cross-reality workflows, including varying object scales, dynamic viewpoint adjustments, precise 2D UI operations (e.g., text editing), and alternative inputs like Gaze+Pinch~\cite{pfeuffer2017gaze} or 6-DoF controllers. Conducting comparative studies across diverse tasks would rigorously benchmark the ecological validity of \papername and provide deeper insights into when users benefit most from continuous multimodal coordination versus hybrid modality switching.

%%%%%%%%%%%%%%%%%%%%%%%%%%%%%%%%%%%%%%%%%%%%%%%%%%%%%%%%%%%%%%%%%%%%%%%%%%%%%%%%%%%%%%%%%%%%%%%%%%

\section{Conclusion}
In this paper, we presented \papername, a multimodal interaction technique designed to bridge the dimensionality gap between standard 2D mouse input and 3D spatial manipulation. By leveraging the spatial alignment of gaze and mouse cursors as a dynamic trigger for plane modulation, our approach eliminates the burden of physical device switching. Our user study systematically explored the foundational design space of \papername, optimizing key variables such as mapping profiles and cursor apertures. We demonstrated that its optimal configuration (overlapXZ mapping with a large gaze cursor) significantly enhances task efficiency and subjective satisfaction compared to the traditional 3D Gizmo interface, while maintaining comparable placement precision. Notably, the strong preference for the overlapXZ mapping confirms that aligning interaction mechanics with users' mental models of intent and action is crucial for designing intuitive spatial translation tasks. Building on these insights, we propose that the gaze-augmented mouse interaction paradigm not only extends the utility of ubiquitous input devices into volumetric spaces but also lays the groundwork for seamless, continuous, and high-precision cross-reality workflows without the need for modality switching. 

%%%%%%%%%%%%%%%%%%%%%%%%%%%%%%%%%%%%%%%%%%%%%%%%%%%%%%%%%%%%%%%%%%%%%%%%%%%%%%%%%%%%%%%%%%%%%%%%%%%%%%%%%%%%%%

\acknowledgments{This work was supported by the National Research Foundation of Korea(NRF) grant funded by the Korea government (MSIT) (RS-2025-00521923).}  

\bibliographystyle{abbrv-doi}

\bibliography{template}
\end{document}